\documentclass[prl,twoside,preprintnumbers,superscriptaddress,twocolumn,nofootinbib,nobibnotes]{revtex4}
\pdfoutput=1
\usepackage{amsmath,graphicx,slashed,multirow,color,setup}
\usepackage[normalem]{ulem}
\usepackage[utf8]{inputenc}

\preprint{NIKHEF 2020-012}
\preprint{ZU-TH 12/20}
\preprint{IPPP/20/13}
\preprint{CERN-TH-2020-071}

\begin{document}
\title{Predictions for $\PZ$-boson production in association with a $b$-jet at $\mathcal{O}(\alpha_s^3)$}

\author{R. Gauld}
\email{r.gauld@nikhef.nl}
\affiliation{Nikhef, Science Park 105, NL-1098 XG Amsterdam, The Netherlands}

\author{A. Gehrmann--De Ridder}
\email{gehra@phys.ethz.ch}
\affiliation{Institute for Theoretical Physics, ETH, CH-8093 Z\"urich, Switzerland}
\affiliation{Department of Physics, University of Z\"urich, CH-8057 Z\"urich, Switzerland}

\author{E.~W.~N.~Glover}
\email{e.w.n.glover@durham.ac.uk}
\affiliation{Institute for Particle Physics Phenomenology, University of Durham, DH1 3LE Durham, United Kingdom}

\author{A. Huss}
\email{alexander.huss@cern.ch}
\affiliation{Institute for Particle Physics Phenomenology, University of Durham, DH1 3LE Durham, United Kingdom}
\affiliation{Theoretical Physics Department, CERN, CH-1211 Geneva 23, Switzerland}

\author{I. Majer}
\email{majeri@phys.ethz.ch}
\affiliation{Institute for Theoretical Physics, ETH, CH-8093 Z\"urich, Switzerland}

\date{\today}

\begin{abstract}
Precise predictions are provided for the production of a $\PZ$-boson and a \bjet in hadron-hadron collisions within the framework of perturbative QCD, at $\mathcal{O}(\alpha_s^3)$. 
To obtain these predictions we perform the first calculation of a hadronic scattering process involving the direct production of a flavoured-jet at next-to-next-to-leading order accuracy in massless QCD, and extend techniques to also account for the impact of finite heavy-quark mass effects.
The predictions are compared to CMS data obtained in $\Pp\Pp$ collisions at a centre-of-mass energy of $8~\TeV$, which are the most precise data from Run I of the LHC for this process, where a good description of the data is achieved. 
To allow this comparison we have performed an unfolding of the data, which overcomes the long-standing issue that the experimental and theoretical definitions of jet flavour are incompatible.
\end{abstract}

\maketitle

\section{Introduction}

The high-energy hadron collisions at the LHC provide an environment in which to search for signals of physics Beyond-the-Standard Model (BSM), both through precision measurements of known processes as well as direct searches. An indispensable tool in this endeavour is the identification of flavoured QCD radiation which is generated in these collisions---e.g. a jet of hadrons which are consistent with being initiated by a quark of a specific flavour (such as a charm or beauty quark).
The LHC collaborations have been successful in identifying such signatures~\cite{Aaij:2015yqa,Aad:2015ydr,Sirunyan:2017ezt}, which has enabled a range of measurements which involve the production of a flavoured jet(s) to be performed.
Crucially, the sensitivity of these measurements (through a comparison to theory) relies on the availability of both precise and consistent predictions for SM processes involving the production of flavoured jets.
However, a number of theoretical and experimental issues currently prohibit such a comparison.

Theoretical predictions for processes involving the direct production of a flavoured-jet in hadron-hadron scattering are currently limited to next-to-leading (\NLO) QCD accuracy interfaced to a Parton Shower (\text{NLO+PS})~\cite{Alioli:2010qp,
Alwall:2014hca,Bellm:2015jjp,Bothmann:2019yzt}.
For the process involving the associated production of a flavoured-jet and $\PZ$-boson (the focus of this work), the theoretical uncertainty of the predicted scattering cross-section is $\approx (8-15)\%$---which is limited by a lack of perturbative knowledge of the scattering process, and often less precise than the available data.
There exists an additional issue for processes involving a \bjet (a jet associated to a $b$-quark), which is that calculations that treat either the $b$-quark as a massive or massless QCD parton can lead to very different results. 
For example, the measured fiducial cross-section for the process $\Pp\Pp \to \PZ+\bjet$ by the CMS collaboration at $8~\TeV$~\cite{Khachatryan:2016iob} indicates that the massive (massless) \text{NLO+PS} calculation predicts a cross section which under (over) estimates the observed one by $\approx 20\%$.

However, the more overarching problem is that the theoretical and experimental definitions of jet flavour are not consistent.
%
Experimentally, measurements involving flavour-tagged jets at the LHC are performed using jets which are reconstructed with the anti-$k_T$ algorithm~\cite{Cacciari:2008gp} and have the property of flavour assigned after the reconstruction process~\cite{Aaij:2015yqa,Aad:2015ydr,Sirunyan:2017ezt}.
The main issue with algorithms of this type is related to how wide-angle soft (low-energy) quark-antiquark pairs are clustered as part of the theoretical prediction. There is a possibility that only one of these (flavoured) soft quarks is clustered into a hard jet, altering its flavour, and thus rendering the definition of jet flavour sensitive to soft physics---such a definition is not infrared safe.
Instead, predictions for flavoured-jet observables provided at fixed-order in perturbative QCD rely on the use of an algorithm that leads to an infrared-safe definition of jet flavour~\cite{Banfi:2006hf}. The application of a flavoured-jet algorithm is necessary for massless computations of this type, which are otherwise not finite.
Consequently, a comparison between a precise massless QCD computation (which, importantly, includes the resummation of initial-state collinear logarithms associated to the flavoured quark) and LHC data is not currently possible.

Overcoming these obstacles is essential to improve the sensitivity of future measurements of the direct production of a flavoured jet(s), which will otherwise be limited to the use of \text{NLO+PS} accurate predictions for comparison.
The goal of this work to address these issues by: performing the first calculation of a hadronic-scattering process involving the direct production of a flavoured jet at next-to-next-to-leading (\NNLO) QCD accuracy; extending the theoretical methods which combine massive and massless QCD calculations; and presenting a solution to allow for a comparison between predictions and data based on an incompatible definitions of jet flavour.

We focus on the process $\Pp\Pp \to \PZ+\bjet$, which is of particular interest due to the high production rate and clean experimental signature. These features have allowed for this process to be measured with high precision~\cite{Khachatryan:2016iob}, which provides an important testing ground for the theoretical developments of this work.
%
In the following, we provide details of the ingredients of the calculation, before providing a comparison to available (unfolded) data from the CMS collaboration in $\Pp\Pp$ collisions at $\sqrt{s} = 8~\TeV$. We conclude with a discussion on the prospects of direct comparisons between perturbative QCD predictions and future LHC measurements.

\section{Details of the Calculation}
In this work we are interested in the prediction of flavoured-jet observables for the process $\Pp\Pp \to \PZ+\bjet$ at $\mathcal{O}(\alpha_s^3)$. We here wish to combine the computation performed in a scheme where the $b$-quark is treated as a massless parton (${\rm 5fs}$) with that where mass effects of the $b$-quark are included exactly (${\rm 4fs}$).
Schematically, the combined cross-section is
\beq \label{eq:sigma}
{\rm d}\sigma^{\rm FONLL} = {\rm d}\sigma^{\rm 5fs} + \left( {\rm d}\sigma_{m_b}^{\rm 4fs} - {\rm d}\sigma^{\rm 4fs}_{m_b\to0} \right)\,,
\eeq
where ${\rm d}\sigma^{\rm 5fs}$ is the massless {\rm 5fs} prediction, ${\rm d}\sigma_{m_b}^{\rm 4fs}$ is the massive {\rm 4fs} prediction,
and ${\rm d}\sigma_{m_b\to0}^{\rm 4fs}$ is the prediction obtained in the massless limit of the {\rm 4fs}. It is further understood that each of these predictions has an expansion in terms of perturbative coefficients.
The computation of all terms in Eq.~\eqref{eq:sigma} will be performed with parton distribution functions (PDFs) and $\alpha_s$ defined in the ${\rm 5fs}$, and as a consequence it is necessary to re-write the contributions in parenthesis in terms of ${\rm 5fs}$ inputs. This can be achieved by applying the relevant $n_f$-dependent scheme corrections to the perturbative coefficients---see for example Eq.~(3.15)--(3.16) of~\cite{Cacciari:1998it}.

The procedure outlined above is often referred to as `FONLL' and has previously been applied to the description of exclusive flavoured hadron final states~\cite{Cacciari:1998it,Cacciari:2001td}, QCD inclusive processes~\cite{Buza:1996wv,Forte:2010ta}, inclusive cross-sections for several processes~\cite{Forte:2015hba,Forte:2016sja,Forte:2018ovl,Duhr:2020kzd}, as well as differential predictions of flavoured jets~\cite{Banfi:2007gu}. An algorithm to apply this method in the context of multijet merging with Parton Showers was also recently developed~\cite{Hoche:2019ncc}. See also~\cite{Collins:1978wz,Aivazis:1993pi,Buza:1996wv,Thorne:1997ga,Kretzer:1998ju,Collins:1998rz,Cacciari:1998it,Kramer:2000hn,Tung:2001mv,Thorne:2006qt,Bierenbaum:2009zt,Forte:2010ta,Guzzi:2011ew,Maltoni:2012pa,Behring:2014eya,Bonvini:2015pxa,Hoang:2015iva,Ablinger:2017err,Krauss:2017wmx,Forte:2019hjc} where these (and alternative) techniques have been developed. Here we extend this work by applying the method to fully differential (flavoured-jet) observables based on a massless \NNLO calculation.

\noindent \textbf{Massless calculation.}
The computation of ${\rm d}\sigma^{\rm 5fs}$ at $\mathcal{O}(\alpha_s^3)$ requires the \NNLO QCD calculation of the process $\Pp\Pp \to \PZ + \bjet$ in the ${\rm 5fs}$.
This has been computed for the first time in this work, based on the calculation of the process $\Pp\Pp \to \PZ + \jet$~\cite{Ridder:2015dxa}.
The computation~\cite{Ridder:2015dxa}, which is agnostic to the flavour of the out-going jet, was performed with the \nnlojet framework~\cite{Ridder:2015dxa} and uses the antenna subtraction method~\cite{GehrmannDeRidder:2005cm,GehrmannDeRidder:2005aw,GehrmannDeRidder:2005hi,Daleo:2006xa,Daleo:2009yj,Boughezal:2010mc,Gehrmann:2011wi,GehrmannDeRidder:2012ja,Currie:2013vh} to obtain fully differential cross-section predictions after the analytical cancellation of all infrared divergences.

The computation of flavour sensitive observables for the $\PZ+\bjet$ process additionally requires the complete flavour and momentum information of all physical (squared) matrix elements and subtraction terms.
This was not available previously in the $\PZ+ \jet$ calculation, but has been incorporated into the \nnlojet framework, allowing for the computation of flavoured-jet observables. See~\cite{Gauld:2019yng} for an overview of this procedure.

\noindent \textbf{Massive calculation.}
To obtain the massive contribution ${\rm d}\sigma_{m_b}^{\rm 4fs}$ at $\mathcal{O}(\alpha_s^3)$, originally computed in~\cite{FebresCordero:2008ci,Cordero:2009kv} for $Z +b \bar{b}$ production at NLO level, we use the automated framework \aMCatNLO~\cite{Hirschi:2011pa,Alwall:2014hca} which has been operated with a number of external libraries~\cite{vanHameren:2009dr,vanHameren:2010cp,Mastrolia:2012bu,Peraro:2014cba,Denner:2016kdg}.

\noindent \textbf{Zero-mass limit.}
When taking the zero-mass limit of the massive coefficient, terms multiplied by power-corrections of the form $m_b^2/Q^2$ vanish, while finite and logarithmically divergent terms without such a pre-factor remain.
These latter contributions are already included within ${\rm d}\sigma^{\rm 5fs}$, and must be subtracted from the massive coefficient to avoid double counting according to Eq.~\eqref{eq:sigma}. 

The finite terms which are present in the zero-mass limit can be obtained from the {\rm 5fs} massless computation discussed above by neglecting all $b$-quark initiated contributions, and by applying the necessary scheme corrections.
The computation of the logarithmically divergent contributions can instead be performed in the following way. First, an expression for the $b$-quark PDF expanded up to $\mathcal{O}(\alpha_s^2)$ using the matching coefficients given in ~\cite{Buza:1996wv} should be obtained. This PDF is then convoluted with the massless partonic cross-section of the massless {\rm 5fs} calculation (also expanded in $\alpha_s$), and the resultant terms of the convolution up to $\mathcal{O}(\alpha_s^3)$ are kept.
The computation of these logarithmic corrections is performed with a specially tailored Monte Carlo programme, which includes the expressions for both the matching coefficients of~\cite{Buza:1996wv} and the massless partonic cross-section up to $\mathcal{O}(\alpha_s^2)$.

\noindent \textbf{Jet algorithm.}
It is essential for fixed-order computations to be applied to observables which are insensitive to both the dynamics of soft and collinear physics.
To this end, we use the flavour-$k_{\rT}$ algorithm originally proposed in~\cite{Banfi:2006hf}.
As compared to standard jet algorithms, the clustering procedure for this algorithm must have both the flavour and momentum information of the input particles. First, the flavour of (pseudo)jets is defined by the net flavour of its constituents, assigning $+1$ ($-1$) if a flavoured quark (antiquark) is present. Second, the definition of the distance measure of this algorithm (which determines the clustering outcome) depends on the flavour of the pseudojet being clustered. These steps are necessary to avoid situations where soft quarks can alter the flavour of a jet, as described above.
In addition, the net flavour criterion also ensures that jets which contain (quasi)collinear quark pairs are not assigned an overall flavour based on such splittings. More details can be found in~\cite{Banfi:2006hf,Banfi:2007gu}.

\section{Comparison with $8~\TeV$ CMS Data}
In this Section we perform a comparison of the $\PZ+\bjet$ CMS data at $8~\TeV$ provided in~\cite{Khachatryan:2016iob}, and validate our implementation of Eq.~\eqref{eq:sigma}. Before doing so we summarise the numerical set-up, and present details on the unfolding procedure which is applied to this data to make a consistent comparison with our theoretical predictions possible.

\noindent \textbf{Numerical inputs.}
All predictions are provided with the NNPDF3.1 \NNLO PDF set~\cite{Ball:2017nwa} with $\alpha_s(M_\PZ) = 0.118$ and $n_f^{\rm max} = 5$, where both the PDF and $\alpha_s$ values are accessed via LHAPDF~\cite{Buckley:2014ana}.
The results are obtained using the $G_{\mu}$-scheme with the following values for the input parameters $M_{\PZ}^\mathrm{os} = 91.1876~\GeV$, $\Gamma_{\PZ}^\mathrm{os} = 2.4952~\GeV$, $M_{\PW}^\mathrm{os} = 80.385~\GeV$,  $\Gamma_{\PW}^\mathrm{os} = 2.085~\GeV,$ and $G_\mu = 1.16638 \times 10^{-5}~\GeV^{-2}$. Including also the universal corrections to the $\rho$ parameter when determining the numerical values of $\alpha$ and $\sin^2\theta_{\text{W}}$ as in~\cite{Gauld:2017tww}, leads to $\alpha_{\rm eff.} = 0.007779$ and $\sin^2\theta_{\text{W,eff.}} = 0.2293$.
An uncertainty due to the impact of missing higher-order corrections is assessed in the predictions by varying the values of $\mu_F$ and $\mu_R$ by a factor of two around the central scale $\mu_0 \equiv E_{\rT,\PZ}$, with the additional constraint that $\frac{1}{2} \leq \mu_F/\mu_R \leq 2$. The scales are treated as correlated between the coefficients appearing in Eq.~\eqref{eq:sigma}.
We follow the specific setup of the flavour-$k_\rT$ algorithm adopted in~\cite{Gauld:2019yng}, where a value of $\alpha = 2$ is used and a beam distance measure that includes a sum over both QCD partons as well as the reconstructed gauge boson is introduced.

\noindent \textbf{Unfolding.} 
As already highlighted, the fixed-order prediction for a flavoured-jet cross-section as defined in Eq.~\eqref{eq:sigma} must be performed with an infrared-safe definition of jet flavour.
However, there is no data available for the process $\Pp\Pp \to \PZ+\bjet$~\cite{Chatrchyan:2012vr,Chatrchyan:2013zja,Chatrchyan:2014dha,Aad:2014dvb,Khachatryan:2016iob,Sirunyan:2020lgh,Aad:2020gfi} (or in fact any process) which uses such a definition of jet flavour. 
%
To address this issue, we have computed a correction to the CMS data~\cite{Khachatryan:2016iob} as described below.

This data has been presented for anti-$k_{\rT}$ ${b\text{-jets}}$, with a flavour assignment based on whether the jet contains $B$-hadron decay products and the additional requirement that $\Delta R(B,\jet) < 0.5$.
To correct this data to the level of partonic flavour-$k_{\rT}$ jets, we apply an unfolding procedure with the \text{RooUnfold}~\cite{Adye:2011gm} package using the iterative Bayes method~\cite{DAgostini:1994fjx}.
The input to this procedure is a theoretical model for the original data using both the anti-$k_\rT$ algorithm (which is measured) and the flavour-$k_\rT$ algorithm (which we wish to unfold to). 

This model is provided with an \text{NLO+PS} prediction for $\PZ+\bjet$ using \aMCatNLO~\cite{Alwall:2014hca} interfaced to \text{Pythia8.243}~\cite{Sjostrand:2014zea}.
%
The parton-level flavour-$k_\rT$ prediction is obtained using the input QCD partons which are identical to those which enter the hadronisation process.
For the central value, we use a {\rm 5fs} prediction of $\PZ+\jet$, where the $\bjet$ contribution of this sample is extracted. 
The benefit of this approach is that the fragmentation component (e.g. $g\to b\bar b$) is resummed by the PS. 
To assess the uncertainty of this procedure, the unfolding is repeated taking into account the impact of scale variations in the model. Additionally, the whole procedure is repeated with a {\rm 4fs} prediction, and the envelope of all of these results is assigned as an uncertainty.
Finally, the unfolding procedure was also performed with a bin-by-bin unfolding method, which led to almost identical results for the considered distributions.

\begin{figure}[h]
  \begin{center}
    \makebox{\includegraphics[width=1.0\columnwidth]{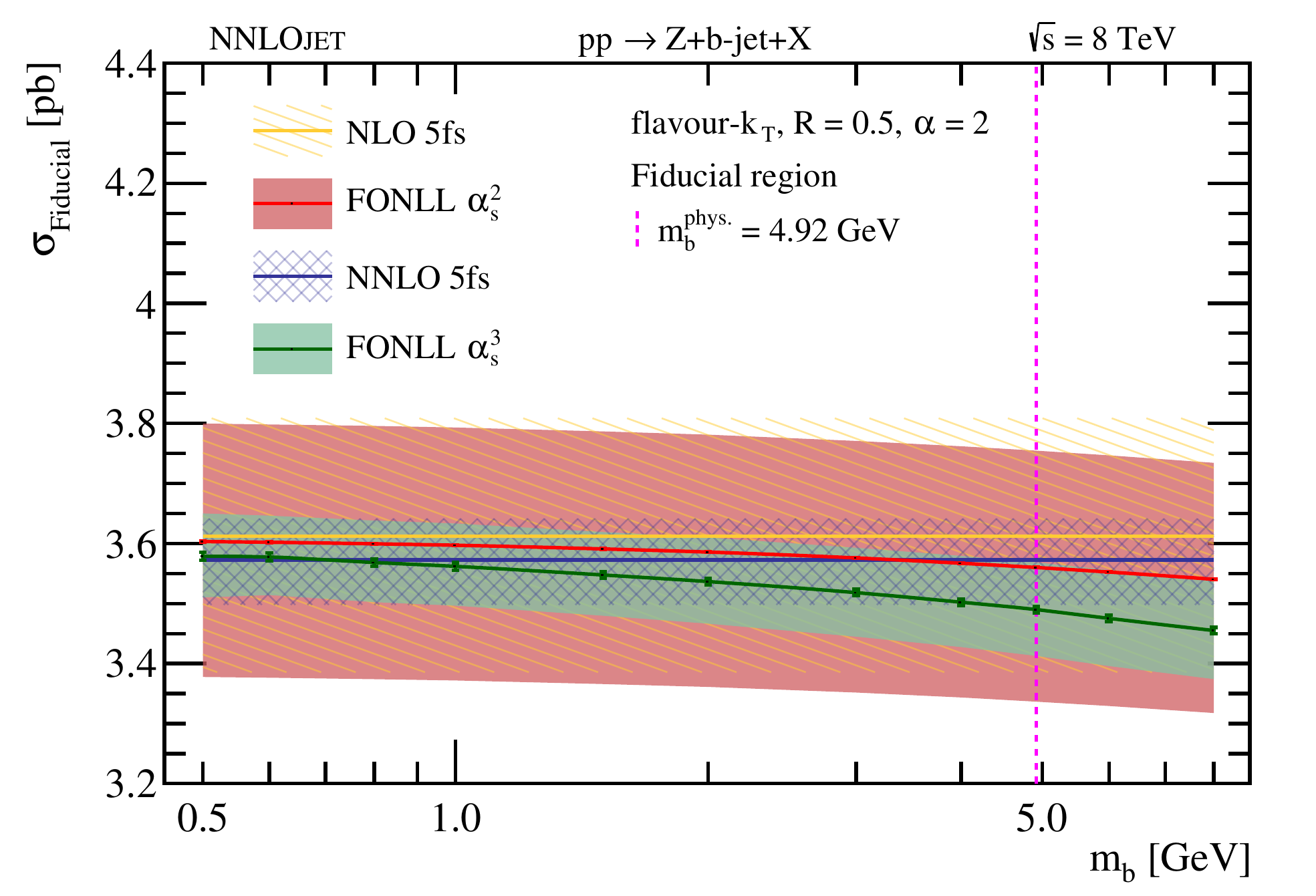}}
  \end{center}
  \vspace{-0.8cm}
  \caption{Fiducial cross-section for the process $\Pp\Pp \to \PZ+\bjet+X$ at $\sqrt{s}=8~\TeV$. The FONLL predictions are provided as a function of $m_b$, and are compared to the {\rm 5fs} predictions.}
  \label{fig:Fiducial}
\end{figure}

\noindent \textbf{Fiducial cross-section.}
In Fig.~\ref{fig:Fiducial}, the cross-section predictions for the process $\Pp\Pp \to \PZ+\bjet$ are shown within the fiducial region defined according to: $p_{\rT,b} > 30~\GeV$, $|\eta_{b}| < 2.4$, $p_{\rT,\ell} > 20~\GeV,$ $|\eta_{\ell}| < 2.4,$ and $M_{\ell\bar{\ell}} \in[71,111]~\GeV$. The \bjets are reconstructed with the flavour-$k_\rT$ algorithm with $R=0.5$, with the additional constraint of $\Delta R(b,\ell) > 0.5$.
As discussed above, this matches the fiducial region of the data~\cite{Khachatryan:2016iob} with the exception of the choice of the jet clustering algorithm.

The cross-section defined according to Eq.~\eqref{eq:sigma} is labelled as `FONLL', and predictions are shown at both $\mathcal{O}(\alphas^2)$ and $\mathcal{O}(\alphas^3)$ as a function of $m_b$ (as it arises explicitly in the parenthesis on the r.h.s. of Eq.~\eqref{eq:sigma}).
The filled band indicates the uncertainty due to scale variation alone, the small error bars on the FONLL predictions indicate numerical uncertainties, and these predictions are then compared to the corresponding {\rm 5fs} scheme predictions at each respective order.
It is found that these two predictions coincide in the limit $m_b\to0$, which demonstrates that both the finite zero-mass and the logarithmically divergent terms have been correctly subtracted from the massive computation, thus providing an important cross-check of our implementation of Eq.~\eqref{eq:sigma}.

The physical prediction is obtained for the $b$-quark mass as indicated by the dashed vertical line at $m^{\rm phys.}_b = 4.92~\GeV$. At $\mathcal{O}(\alpha_s^3)$, the FONLL prediction is $\sigma^{\rm FONLL}_{\rm Fiducial}(m^{\rm phys.}_b) = 3.490^{+0.078}_{-0.078}{\rm (scales)}~\pb$. As compared to $\mathcal{O}(\alpha_s^2)$, a large reduction in the scale uncertainty of the prediction and a small negative shift on the central value is observed.
Furthermore, it is found that the inclusion of mass corrections at $\mathcal{O}(\alpha_s^3)$ leads to a negative correction ($-2.3\%$). The impact of the mass corrections is as large as the scale uncertainty, which underpins the importance of including such corrections as part of a precision computation.

To compare this prediction to data, we perform the unfolding procedure for the fiducial cross-section region defined in~\cite{Khachatryan:2016iob}, finding a correction of $c = 0.883^{+0.004}_{-0.008}$.
It is found that the main contribution to this correction is the subtraction of a `fake' rate from the data, corresponding to situations where an event which passes the fiducial selection when the anti-$k_\rT$ clustering is used, but does not pass the same selection when instead the flavour-$k_\rT$ clustering is employed.
Applying this correction to the data gives $\sigma_{\rm Fiducial,f\text{-}k_\rT}^{\rm CMS} = 3.134\pm0.214^{+0.013}_{-0.025}\pb$, where the first uncertainty is that of the original measurement and the second one due to the unfolding procedure.
With respect to the central value of the FONLL $\mathcal{O}(\alpha_s^3)$ prediction, taking only the experimental uncertainty into account, the agreement with the unfolded data is $1.67\sigma$.
In addition to the scale uncertainty shown in Fig.~\ref{fig:Fiducial}, an uncertainty due to PDF and variation of $\alpha_s(M_{\PZ}) = 0.118\pm0.001$ has also been assessed (at \NLO), which gives $\delta \sigma({\rm PDF,\alpha_s}) = \pm0.074~\pb$. The uncertainty of the prediction and unfolded data overlap when these additional sources of uncertainty are taken into account.

\noindent \textbf{Differential distributions.}
As part of the measurement~\cite{Khachatryan:2016iob}, a number of differential observables for the process $\Pp\Pp \to \PZ+\bjet$ were considered.
Here we have chosen to focus on the transverse momentum of the leading \bjet ($p_{\rT,b}$) as well as the absolute pseudorapidity of the leading \bjet ($\eta_b$). 

\begin{figure}[h]
  \begin{center}
    \makebox{\includegraphics[width=1.0\columnwidth]{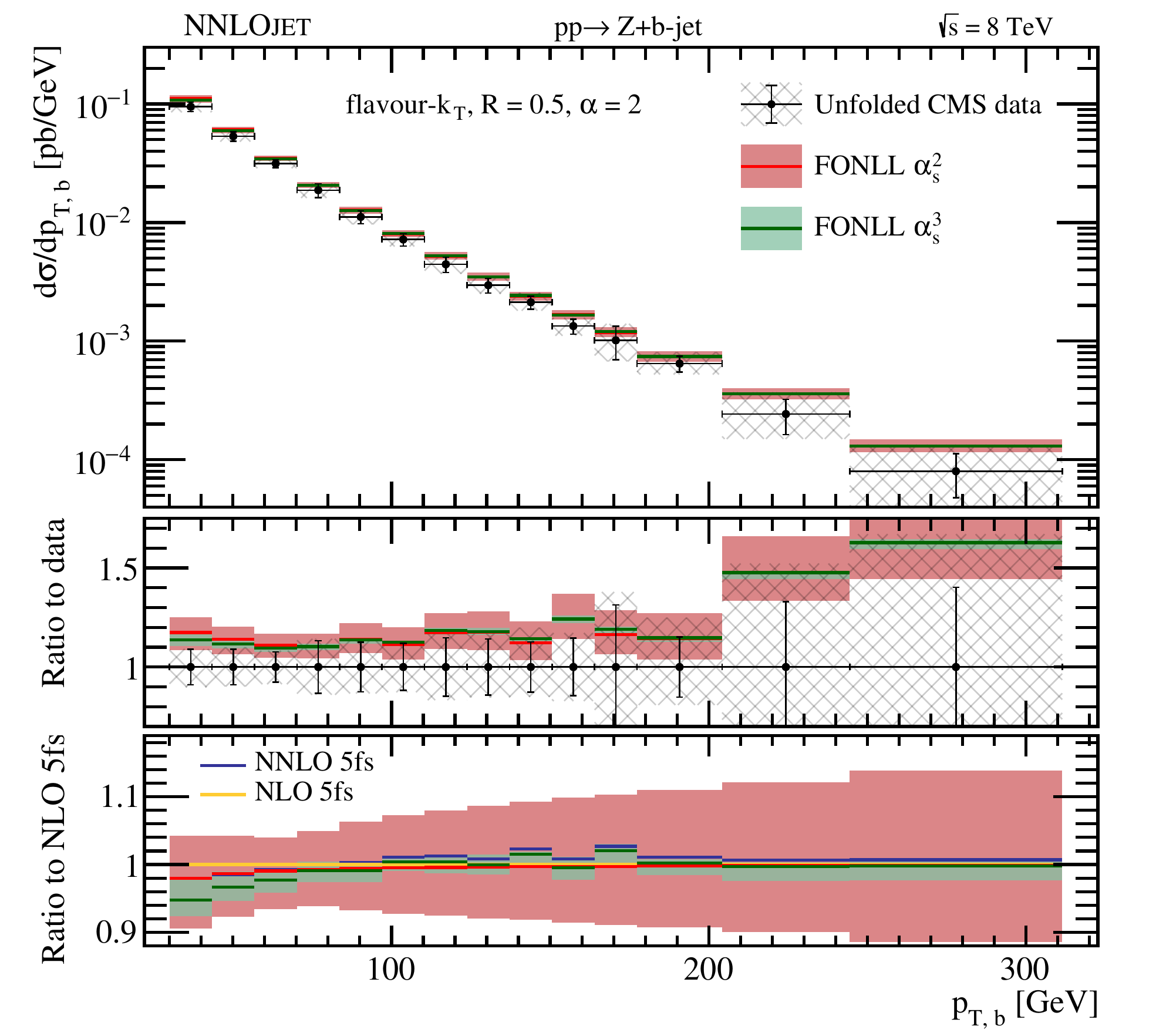}}
  \end{center}
  \vspace{-0.8cm}
  \caption{The transverse momentum distribution of the leading flavour-$k_{\rT}$ \bjet. The absolute cross-section is shown in the upper panel, the ratio to the unfolded data in the central panel, and the ratio to the NLO {\rm 5fs} prediction in the lower panel. The shown uncertainty of the FONLL distributions are due to scale variations alone.}
  \label{fig:dptj}
\end{figure}

The $p_{\rT,b}$ distribution is shown in Fig.~\ref{fig:dptj} where the absolute cross-section is shown in the upper panel, the ratio to data in the central panel, and the ratio to the \NLO~{\rm 5fs} prediction in the lower panel. The FONLL predictions are provided at the physical $b$-quark mass, and the uncertainty due to scale variation is shown. The central result of the unfolded CMS data is indicated with black error bars, and the additional uncertainty due to the input model of the unfolding procedure is overlaid with a grey crossed fill. In the lower panel, we have included the central (N)NLO predictions in the {\rm 5fs} scheme to indicate the relevance of the mass corrections.
A large reduction in the scale uncertainties for this distribution are observed at $\mathcal{O}(\alpha_s^3)$. The impact of the mass corrections is most relevant at small values of $p_{\rT,b}$, where they approximately amount to $-4\%$, while for large $p_{\rT,b}$ they essentially vanish. This behaviour is naively expected as a scale set by the power corrections is of the form $m_b^2/p_{\rT,b}^2$.
Reasonable agreement with the data is found, although there is a tendency for the data to prefer a smaller normalisation.
To better quantify this agreement, we have computed the $\chi^2$ for this observable with respect to the central FONLL predictions, finding $\chi^2/{\rm N_{dat}}(\alpha_s^2, p_{\rT,b}) = 23.4/14$ and $\chi^2/{\rm N_{dat}}(\alpha_s^3,p_{\rT,b}) = 21.5/14$.
This is an underestimate of the agreement as no correlations have been included in this test---they are not publicly available---and only the experimental (inner) uncertainty of the unfolded data has been used. 

\begin{figure}[h]
  \begin{center}
    \makebox{\includegraphics[width=1.0\columnwidth]{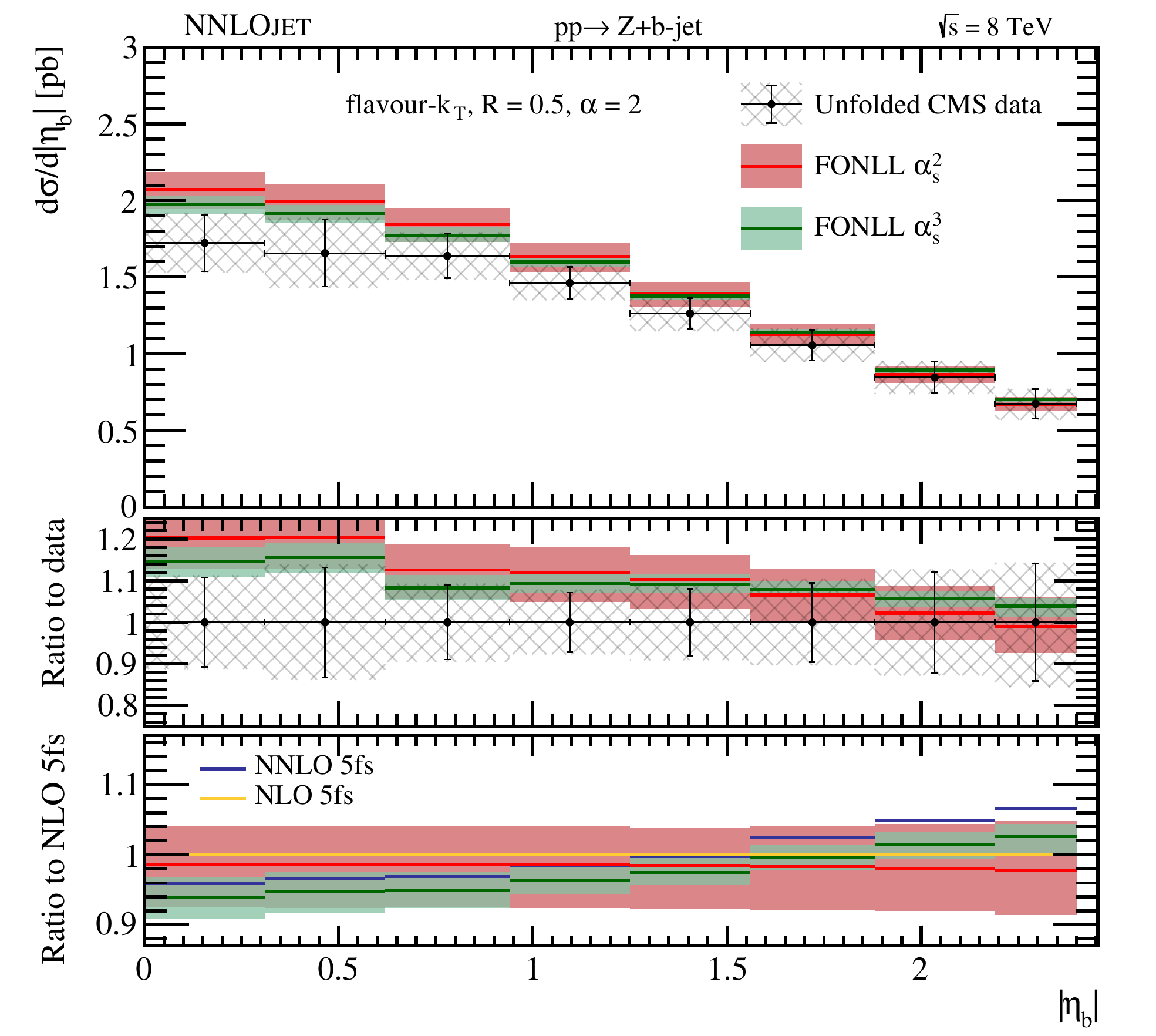}}
  \end{center}
  \vspace{-0.8cm}
  \caption{As in Fig.~\ref{fig:dptj}, now for the absolute psudorapidity distribution of the leading flavour-$k_{\rT}$ \bjet.}
  \label{fig:detaj}
\end{figure}

The corresponding Figure for the $|\eta_b|$ distribution is shown in Fig.~\ref{fig:detaj}. As before, the $\mathcal{O}(\alpha_s^3)$ corrections are essential for improving the precision of the theory predictions. These mass corrections are negative, and range from $-2\%$ at central pseudrapidities to $-4\%$ in the forward region. The mass corrections are observed to be most important for the $q\bar q$-induced channel, and therefore become more important at larger pseudorapidity values where the relative contribution of this channel increases.
These corrections are important for improving the description of the data, particularly at central pseudorapidity values where the absolute cross-section is largest. Performing the chi-squared test as above leads to $\chi^2/{\rm N_{dat}}(\alpha_s^2,\eta_b) = 12.9/8$ and $\chi^2/{\rm N_{dat}}(\alpha_s^3,\eta_b) = 8.08/8$, therefore finding agreement between the most precise theoretical prediction and the unfolded data.

\section{Discussion and Conclusions}
In this work, we have performed a precision calculation for observables related to the process $\Pp\Pp \to \PZ+\bjet$. This has been achieved by combining a massless \NNLO and a massive \NLO computations at $\mathcal{O}(\alpha_s^3)$. This is the first time that such a matching has been performed with a fully differential \NNLO massless computation. The predictions exhibit greatly reduced uncertainties and open the door for precision studies involving flavoured jets.
The benefit of this approach is that the contribution to the cross-section which arises from collinear initial-state splittings of the form $g\to b\bar b$, can be conveniently resummed by PDF evolution as part of the massless calculation. This approach is suitable for all processes where these type of logarithmic corrections dominate the cross-section. At the same time, the impact of finite $b$-quark mass effects can easily be incorporated.
As a consequence of using a massless calculation, it becomes necessary to use an infrared-safe definition of jet flavour, which does not align with the current choice made by experimentalists.

To tackle this issue, we have taken the approach to unfold the experimental data which allows for a consistent comparison between the precise theoretical computation with data. We have found reasonable agreement for the leading-$\bjet$ $p_{T,b}$ and $\eta_b$ distributions, as well as the integrated cross-section.
However, a more direct comparison could be possible if the data were directly unfolded to the level of partonic flavour-$k_{\rT}$ jets by the experimental collaborations. This is likely possible as these measurements, such as~\cite{Khachatryan:2016iob}, are already unfolded to a stable particle level to account for event selection efficiencies as well as detector resolution effects. This more direct approach could potentially avoid systematic uncertainties introduced by performing the unfolding twice.
An alternative approach would be for the measurement to be directly performed with flavour-$k_{\rT}$ jets. To our knowledge, there have been no experimental studies which attempt to include flavour information during the jet reconstruction, and so it is not clear how feasible an experimental realisation of the flavour-$k_{\rT}$ algorithm will be.

It is our advice that each of these approaches receive further investigation. In addition to the final states with $\bjets$, charm tagged flavour-$k_{\rT}$ jets should also be considered. This is of relevance for final states such as $\PW/\PZ+\cjet$, where a precise comparison between theory and data is highly desirable, to enable the extraction of the flavour structure of the proton~\cite{Chatrchyan:2013mza,Aad:2014xca,Alekhin:2014sya,Alekhin:2017olj,Boettcher:2015sqn}.

{\it Acknowledgements.}
We are grateful to P. Ilten, V. Hirschi, D. Napoletano, and G. Salam for useful comments/suggestions on various aspects of this work. We are also appreciative of the private implementation of the flavour-$k_{\rT}$ algorithm provided by G. Salam.
The authors also thank Xuan Chen, Juan Cruz-Martinez, James Currie, Thomas Gehrmann, Marius H\"ofer, Jonathan Mo, Tom Morgan, Jan Niehues, Joao Pires, Duncan Walker, and James Whitehead for useful discussions and their many contributions to the \nnlojet code.
This research is supported by the Dutch Organisation for Scientific Research (NWO) through the VENI grant 680-47-461 and by the Swiss National Science Foundation (SNF) under contract 200021-172478.

\bibliography{flav}

\begin{thebibliography}{73}
\expandafter\ifx\csname natexlab\endcsname\relax\def\natexlab#1{#1}\fi
\expandafter\ifx\csname bibnamefont\endcsname\relax
  \def\bibnamefont#1{#1}\fi
\expandafter\ifx\csname bibfnamefont\endcsname\relax
  \def\bibfnamefont#1{#1}\fi
\expandafter\ifx\csname citenamefont\endcsname\relax
  \def\citenamefont#1{#1}\fi
\expandafter\ifx\csname url\endcsname\relax
  \def\url#1{\texttt{#1}}\fi
\expandafter\ifx\csname urlprefix\endcsname\relax\def\urlprefix{URL }\fi
\providecommand{\bibinfo}[2]{#2}
\providecommand{\eprint}[2][]{\url{#2}}

\bibitem[{\citenamefont{Aaij et~al.}(2015)}]{Aaij:2015yqa}
\bibinfo{author}{\bibfnamefont{R.}~\bibnamefont{Aaij}} \bibnamefont{et~al.}
  (\bibinfo{collaboration}{LHCb}), \bibinfo{journal}{JINST}
  \textbf{\bibinfo{volume}{10}}, \bibinfo{pages}{P06013}
  (\bibinfo{year}{2015}), \eprint{1504.07670}.

\bibitem[{\citenamefont{Aad et~al.}(2016)}]{Aad:2015ydr}
\bibinfo{author}{\bibfnamefont{G.}~\bibnamefont{Aad}} \bibnamefont{et~al.}
  (\bibinfo{collaboration}{ATLAS}), \bibinfo{journal}{JINST}
  \textbf{\bibinfo{volume}{11}}, \bibinfo{pages}{P04008}
  (\bibinfo{year}{2016}), \eprint{1512.01094}.

\bibitem[{\citenamefont{Sirunyan et~al.}(2018)}]{Sirunyan:2017ezt}
\bibinfo{author}{\bibfnamefont{A.~M.} \bibnamefont{Sirunyan}}
  \bibnamefont{et~al.} (\bibinfo{collaboration}{CMS}), \bibinfo{journal}{JINST}
  \textbf{\bibinfo{volume}{13}}, \bibinfo{pages}{P05011}
  (\bibinfo{year}{2018}), \eprint{1712.07158}.

\bibitem[{\citenamefont{Alioli et~al.}(2011)\citenamefont{Alioli, Nason,
  Oleari, and Re}}]{Alioli:2010qp}
\bibinfo{author}{\bibfnamefont{S.}~\bibnamefont{Alioli}},
  \bibinfo{author}{\bibfnamefont{P.}~\bibnamefont{Nason}},
  \bibinfo{author}{\bibfnamefont{C.}~\bibnamefont{Oleari}}, \bibnamefont{and}
  \bibinfo{author}{\bibfnamefont{E.}~\bibnamefont{Re}}, \bibinfo{journal}{JHEP}
  \textbf{\bibinfo{volume}{01}}, \bibinfo{pages}{095} (\bibinfo{year}{2011}),
  \eprint{1009.5594}.

\bibitem[{\citenamefont{Alwall et~al.}(2014)\citenamefont{Alwall, Frederix,
  Frixione, Hirschi, Maltoni et~al.}}]{Alwall:2014hca}
\bibinfo{author}{\bibfnamefont{J.}~\bibnamefont{Alwall}},
  \bibinfo{author}{\bibfnamefont{R.}~\bibnamefont{Frederix}},
  \bibinfo{author}{\bibfnamefont{S.}~\bibnamefont{Frixione}},
  \bibinfo{author}{\bibfnamefont{V.}~\bibnamefont{Hirschi}},
  \bibinfo{author}{\bibfnamefont{F.}~\bibnamefont{Maltoni}},
  \bibnamefont{et~al.}, \bibinfo{journal}{JHEP}
  \textbf{\bibinfo{volume}{1407}}, \bibinfo{pages}{079} (\bibinfo{year}{2014}),
  \eprint{1405.0301}.

\bibitem[{\citenamefont{Bellm et~al.}(2016)}]{Bellm:2015jjp}
\bibinfo{author}{\bibfnamefont{J.}~\bibnamefont{Bellm}} \bibnamefont{et~al.},
  \bibinfo{journal}{Eur. Phys. J.} \textbf{\bibinfo{volume}{C76}},
  \bibinfo{pages}{196} (\bibinfo{year}{2016}), \eprint{1512.01178}.

\bibitem[{\citenamefont{Bothmann et~al.}(2019)}]{Bothmann:2019yzt}
\bibinfo{author}{\bibfnamefont{E.}~\bibnamefont{Bothmann}} \bibnamefont{et~al.}
  (\bibinfo{collaboration}{Sherpa}), \bibinfo{journal}{SciPost Phys.}
  \textbf{\bibinfo{volume}{7}}, \bibinfo{pages}{034} (\bibinfo{year}{2019}),
  \eprint{1905.09127}.

\bibitem[{\citenamefont{Khachatryan et~al.}(2017)}]{Khachatryan:2016iob}
\bibinfo{author}{\bibfnamefont{V.}~\bibnamefont{Khachatryan}}
  \bibnamefont{et~al.} (\bibinfo{collaboration}{CMS}), \bibinfo{journal}{Eur.
  Phys. J.} \textbf{\bibinfo{volume}{C77}}, \bibinfo{pages}{751}
  (\bibinfo{year}{2017}), \eprint{1611.06507}.

\bibitem[{\citenamefont{Cacciari et~al.}(2008)\citenamefont{Cacciari, Salam,
  and Soyez}}]{Cacciari:2008gp}
\bibinfo{author}{\bibfnamefont{M.}~\bibnamefont{Cacciari}},
  \bibinfo{author}{\bibfnamefont{G.~P.} \bibnamefont{Salam}}, \bibnamefont{and}
  \bibinfo{author}{\bibfnamefont{G.}~\bibnamefont{Soyez}},
  \bibinfo{journal}{JHEP} \textbf{\bibinfo{volume}{0804}}, \bibinfo{pages}{063}
  (\bibinfo{year}{2008}), \eprint{0802.1189}.

\bibitem[{\citenamefont{Banfi et~al.}(2006)\citenamefont{Banfi, Salam, and
  Zanderighi}}]{Banfi:2006hf}
\bibinfo{author}{\bibfnamefont{A.}~\bibnamefont{Banfi}},
  \bibinfo{author}{\bibfnamefont{G.~P.} \bibnamefont{Salam}}, \bibnamefont{and}
  \bibinfo{author}{\bibfnamefont{G.}~\bibnamefont{Zanderighi}},
  \bibinfo{journal}{Eur. Phys. J.} \textbf{\bibinfo{volume}{C47}},
  \bibinfo{pages}{113} (\bibinfo{year}{2006}), \eprint{hep-ph/0601139}.

\bibitem[{\citenamefont{Cacciari et~al.}(1998)\citenamefont{Cacciari, Greco,
  and Nason}}]{Cacciari:1998it}
\bibinfo{author}{\bibfnamefont{M.}~\bibnamefont{Cacciari}},
  \bibinfo{author}{\bibfnamefont{M.}~\bibnamefont{Greco}}, \bibnamefont{and}
  \bibinfo{author}{\bibfnamefont{P.}~\bibnamefont{Nason}},
  \bibinfo{journal}{JHEP} \textbf{\bibinfo{volume}{9805}}, \bibinfo{pages}{007}
  (\bibinfo{year}{1998}), \eprint{hep-ph/9803400}.

\bibitem[{\citenamefont{Cacciari et~al.}(2001)\citenamefont{Cacciari, Frixione,
  and Nason}}]{Cacciari:2001td}
\bibinfo{author}{\bibfnamefont{M.}~\bibnamefont{Cacciari}},
  \bibinfo{author}{\bibfnamefont{S.}~\bibnamefont{Frixione}}, \bibnamefont{and}
  \bibinfo{author}{\bibfnamefont{P.}~\bibnamefont{Nason}},
  \bibinfo{journal}{JHEP} \textbf{\bibinfo{volume}{0103}}, \bibinfo{pages}{006}
  (\bibinfo{year}{2001}), \eprint{hep-ph/0102134}.

\bibitem[{\citenamefont{Buza et~al.}(1998)\citenamefont{Buza, Matiounine,
  Smith, and van Neerven}}]{Buza:1996wv}
\bibinfo{author}{\bibfnamefont{M.}~\bibnamefont{Buza}},
  \bibinfo{author}{\bibfnamefont{Y.}~\bibnamefont{Matiounine}},
  \bibinfo{author}{\bibfnamefont{J.}~\bibnamefont{Smith}}, \bibnamefont{and}
  \bibinfo{author}{\bibfnamefont{W.~L.} \bibnamefont{van Neerven}},
  \bibinfo{journal}{Eur. Phys. J.} \textbf{\bibinfo{volume}{C1}},
  \bibinfo{pages}{301} (\bibinfo{year}{1998}), \eprint{hep-ph/9612398}.

\bibitem[{\citenamefont{Forte et~al.}(2010)\citenamefont{Forte, Laenen, Nason,
  and Rojo}}]{Forte:2010ta}
\bibinfo{author}{\bibfnamefont{S.}~\bibnamefont{Forte}},
  \bibinfo{author}{\bibfnamefont{E.}~\bibnamefont{Laenen}},
  \bibinfo{author}{\bibfnamefont{P.}~\bibnamefont{Nason}}, \bibnamefont{and}
  \bibinfo{author}{\bibfnamefont{J.}~\bibnamefont{Rojo}},
  \bibinfo{journal}{Nucl. Phys.} \textbf{\bibinfo{volume}{B834}},
  \bibinfo{pages}{116} (\bibinfo{year}{2010}), \eprint{1001.2312}.

\bibitem[{\citenamefont{Forte et~al.}(2015)\citenamefont{Forte, Napoletano, and
  Ubiali}}]{Forte:2015hba}
\bibinfo{author}{\bibfnamefont{S.}~\bibnamefont{Forte}},
  \bibinfo{author}{\bibfnamefont{D.}~\bibnamefont{Napoletano}},
  \bibnamefont{and} \bibinfo{author}{\bibfnamefont{M.}~\bibnamefont{Ubiali}},
  \bibinfo{journal}{Phys. Lett.} \textbf{\bibinfo{volume}{B751}},
  \bibinfo{pages}{331} (\bibinfo{year}{2015}), \eprint{1508.01529}.

\bibitem[{\citenamefont{Forte et~al.}(2016)\citenamefont{Forte, Napoletano, and
  Ubiali}}]{Forte:2016sja}
\bibinfo{author}{\bibfnamefont{S.}~\bibnamefont{Forte}},
  \bibinfo{author}{\bibfnamefont{D.}~\bibnamefont{Napoletano}},
  \bibnamefont{and} \bibinfo{author}{\bibfnamefont{M.}~\bibnamefont{Ubiali}},
  \bibinfo{journal}{Phys. Lett.} \textbf{\bibinfo{volume}{B763}},
  \bibinfo{pages}{190} (\bibinfo{year}{2016}), \eprint{1607.00389}.

\bibitem[{\citenamefont{Forte et~al.}(2018)\citenamefont{Forte, Napoletano, and
  Ubiali}}]{Forte:2018ovl}
\bibinfo{author}{\bibfnamefont{S.}~\bibnamefont{Forte}},
  \bibinfo{author}{\bibfnamefont{D.}~\bibnamefont{Napoletano}},
  \bibnamefont{and} \bibinfo{author}{\bibfnamefont{M.}~\bibnamefont{Ubiali}},
  \bibinfo{journal}{Eur. Phys. J.} \textbf{\bibinfo{volume}{C78}},
  \bibinfo{pages}{932} (\bibinfo{year}{2018}), \eprint{1803.10248}.

\bibitem[{\citenamefont{Duhr et~al.}(2020)\citenamefont{Duhr, Dulat, Hirschi,
  and Mistlberger}}]{Duhr:2020kzd}
\bibinfo{author}{\bibfnamefont{C.}~\bibnamefont{Duhr}},
  \bibinfo{author}{\bibfnamefont{F.}~\bibnamefont{Dulat}},
  \bibinfo{author}{\bibfnamefont{V.}~\bibnamefont{Hirschi}}, \bibnamefont{and}
  \bibinfo{author}{\bibfnamefont{B.}~\bibnamefont{Mistlberger}}
  (\bibinfo{year}{2020}), \eprint{2004.04752}.

\bibitem[{\citenamefont{Banfi et~al.}(2007)\citenamefont{Banfi, Salam, and
  Zanderighi}}]{Banfi:2007gu}
\bibinfo{author}{\bibfnamefont{A.}~\bibnamefont{Banfi}},
  \bibinfo{author}{\bibfnamefont{G.~P.} \bibnamefont{Salam}}, \bibnamefont{and}
  \bibinfo{author}{\bibfnamefont{G.}~\bibnamefont{Zanderighi}},
  \bibinfo{journal}{JHEP} \textbf{\bibinfo{volume}{07}}, \bibinfo{pages}{026}
  (\bibinfo{year}{2007}), \eprint{0704.2999}.

\bibitem[{\citenamefont{Hoche et~al.}(2019)\citenamefont{Hoche, Krause, and
  Siegert}}]{Hoche:2019ncc}
\bibinfo{author}{\bibfnamefont{S.}~\bibnamefont{Hoche}},
  \bibinfo{author}{\bibfnamefont{J.}~\bibnamefont{Krause}}, \bibnamefont{and}
  \bibinfo{author}{\bibfnamefont{F.}~\bibnamefont{Siegert}},
  \bibinfo{journal}{Phys. Rev.} \textbf{\bibinfo{volume}{D100}},
  \bibinfo{pages}{014011} (\bibinfo{year}{2019}), \eprint{1904.09382}.

\bibitem[{\citenamefont{Collins et~al.}(1978)\citenamefont{Collins, Wilczek,
  and Zee}}]{Collins:1978wz}
\bibinfo{author}{\bibfnamefont{J.~C.} \bibnamefont{Collins}},
  \bibinfo{author}{\bibfnamefont{F.}~\bibnamefont{Wilczek}}, \bibnamefont{and}
  \bibinfo{author}{\bibfnamefont{A.}~\bibnamefont{Zee}},
  \bibinfo{journal}{Phys. Rev.} \textbf{\bibinfo{volume}{D18}},
  \bibinfo{pages}{242} (\bibinfo{year}{1978}).

\bibitem[{\citenamefont{Aivazis et~al.}(1994)\citenamefont{Aivazis, Collins,
  Olness, and Tung}}]{Aivazis:1993pi}
\bibinfo{author}{\bibfnamefont{M.~A.~G.} \bibnamefont{Aivazis}},
  \bibinfo{author}{\bibfnamefont{J.~C.} \bibnamefont{Collins}},
  \bibinfo{author}{\bibfnamefont{F.~I.} \bibnamefont{Olness}},
  \bibnamefont{and} \bibinfo{author}{\bibfnamefont{W.-K.} \bibnamefont{Tung}},
  \bibinfo{journal}{Phys. Rev.} \textbf{\bibinfo{volume}{D50}},
  \bibinfo{pages}{3102} (\bibinfo{year}{1994}), \eprint{hep-ph/9312319}.

\bibitem[{\citenamefont{Thorne and Roberts}(1998)}]{Thorne:1997ga}
\bibinfo{author}{\bibfnamefont{R.~S.} \bibnamefont{Thorne}} \bibnamefont{and}
  \bibinfo{author}{\bibfnamefont{R.~G.} \bibnamefont{Roberts}},
  \bibinfo{journal}{Phys. Rev.} \textbf{\bibinfo{volume}{D57}},
  \bibinfo{pages}{6871} (\bibinfo{year}{1998}), \eprint{hep-ph/9709442}.

\bibitem[{\citenamefont{Kretzer and Schienbein}(1998)}]{Kretzer:1998ju}
\bibinfo{author}{\bibfnamefont{S.}~\bibnamefont{Kretzer}} \bibnamefont{and}
  \bibinfo{author}{\bibfnamefont{I.}~\bibnamefont{Schienbein}},
  \bibinfo{journal}{Phys. Rev.} \textbf{\bibinfo{volume}{D58}},
  \bibinfo{pages}{094035} (\bibinfo{year}{1998}), \eprint{hep-ph/9805233}.

\bibitem[{\citenamefont{Collins}(1998)}]{Collins:1998rz}
\bibinfo{author}{\bibfnamefont{J.~C.} \bibnamefont{Collins}},
  \bibinfo{journal}{Phys. Rev.} \textbf{\bibinfo{volume}{D58}},
  \bibinfo{pages}{094002} (\bibinfo{year}{1998}), \eprint{hep-ph/9806259}.

\bibitem[{\citenamefont{Kramer et~al.}(2000)\citenamefont{Kramer, Olness, and
  Soper}}]{Kramer:2000hn}
\bibinfo{author}{\bibfnamefont{M.}~\bibnamefont{Kramer}},
  \bibinfo{author}{\bibfnamefont{F.~I.} \bibnamefont{Olness}},
  \bibnamefont{and} \bibinfo{author}{\bibfnamefont{D.~E.} \bibnamefont{Soper}},
  \bibinfo{journal}{Phys. Rev.} \textbf{\bibinfo{volume}{D62}},
  \bibinfo{pages}{096007} (\bibinfo{year}{2000}), \eprint{hep-ph/0003035}.

\bibitem[{\citenamefont{Tung et~al.}(2002)\citenamefont{Tung, Kretzer, and
  Schmidt}}]{Tung:2001mv}
\bibinfo{author}{\bibfnamefont{W.-K.} \bibnamefont{Tung}},
  \bibinfo{author}{\bibfnamefont{S.}~\bibnamefont{Kretzer}}, \bibnamefont{and}
  \bibinfo{author}{\bibfnamefont{C.}~\bibnamefont{Schmidt}},
  \bibinfo{journal}{J. Phys.} \textbf{\bibinfo{volume}{G28}},
  \bibinfo{pages}{983} (\bibinfo{year}{2002}), \eprint{hep-ph/0110247}.

\bibitem[{\citenamefont{Thorne}(2006)}]{Thorne:2006qt}
\bibinfo{author}{\bibfnamefont{R.~S.} \bibnamefont{Thorne}},
  \bibinfo{journal}{Phys. Rev.} \textbf{\bibinfo{volume}{D73}},
  \bibinfo{pages}{054019} (\bibinfo{year}{2006}), \eprint{hep-ph/0601245}.

\bibitem[{\citenamefont{Bierenbaum et~al.}(2009)\citenamefont{Bierenbaum,
  Blumlein, and Klein}}]{Bierenbaum:2009zt}
\bibinfo{author}{\bibfnamefont{I.}~\bibnamefont{Bierenbaum}},
  \bibinfo{author}{\bibfnamefont{J.}~\bibnamefont{Blumlein}}, \bibnamefont{and}
  \bibinfo{author}{\bibfnamefont{S.}~\bibnamefont{Klein}},
  \bibinfo{journal}{Phys. Lett. B} \textbf{\bibinfo{volume}{672}},
  \bibinfo{pages}{401} (\bibinfo{year}{2009}), \eprint{0901.0669}.

\bibitem[{\citenamefont{Guzzi et~al.}(2012)\citenamefont{Guzzi, Nadolsky, Lai,
  and Yuan}}]{Guzzi:2011ew}
\bibinfo{author}{\bibfnamefont{M.}~\bibnamefont{Guzzi}},
  \bibinfo{author}{\bibfnamefont{P.~M.} \bibnamefont{Nadolsky}},
  \bibinfo{author}{\bibfnamefont{H.-L.} \bibnamefont{Lai}}, \bibnamefont{and}
  \bibinfo{author}{\bibfnamefont{C.-P.} \bibnamefont{Yuan}},
  \bibinfo{journal}{Phys.Rev.} \textbf{\bibinfo{volume}{D86}},
  \bibinfo{pages}{053005} (\bibinfo{year}{2012}), \eprint{1108.5112}.

\bibitem[{\citenamefont{Maltoni et~al.}(2012)\citenamefont{Maltoni, Ridolfi,
  and Ubiali}}]{Maltoni:2012pa}
\bibinfo{author}{\bibfnamefont{F.}~\bibnamefont{Maltoni}},
  \bibinfo{author}{\bibfnamefont{G.}~\bibnamefont{Ridolfi}}, \bibnamefont{and}
  \bibinfo{author}{\bibfnamefont{M.}~\bibnamefont{Ubiali}},
  \bibinfo{journal}{JHEP} \textbf{\bibinfo{volume}{07}}, \bibinfo{pages}{022}
  (\bibinfo{year}{2012}), \bibinfo{note}{[Erratum: JHEP04,095(2013)]},
  \eprint{1203.6393}.

\bibitem[{\citenamefont{Behring et~al.}(2014)\citenamefont{Behring, Bierenbaum,
  Blümlein, De~Freitas, Klein, and Wißbrock}}]{Behring:2014eya}
\bibinfo{author}{\bibfnamefont{A.}~\bibnamefont{Behring}},
  \bibinfo{author}{\bibfnamefont{I.}~\bibnamefont{Bierenbaum}},
  \bibinfo{author}{\bibfnamefont{J.}~\bibnamefont{Blümlein}},
  \bibinfo{author}{\bibfnamefont{A.}~\bibnamefont{De~Freitas}},
  \bibinfo{author}{\bibfnamefont{S.}~\bibnamefont{Klein}}, \bibnamefont{and}
  \bibinfo{author}{\bibfnamefont{F.}~\bibnamefont{Wißbrock}},
  \bibinfo{journal}{Eur. Phys. J. C} \textbf{\bibinfo{volume}{74}},
  \bibinfo{pages}{3033} (\bibinfo{year}{2014}), \eprint{1403.6356}.

\bibitem[{\citenamefont{Bonvini et~al.}(2015)\citenamefont{Bonvini,
  Papanastasiou, and Tackmann}}]{Bonvini:2015pxa}
\bibinfo{author}{\bibfnamefont{M.}~\bibnamefont{Bonvini}},
  \bibinfo{author}{\bibfnamefont{A.~S.} \bibnamefont{Papanastasiou}},
  \bibnamefont{and} \bibinfo{author}{\bibfnamefont{F.~J.}
  \bibnamefont{Tackmann}}, \bibinfo{journal}{JHEP}
  \textbf{\bibinfo{volume}{11}}, \bibinfo{pages}{196} (\bibinfo{year}{2015}),
  \eprint{1508.03288}.

\bibitem[{\citenamefont{Hoang et~al.}(2016)\citenamefont{Hoang, Pietrulewicz,
  and Samitz}}]{Hoang:2015iva}
\bibinfo{author}{\bibfnamefont{A.~H.} \bibnamefont{Hoang}},
  \bibinfo{author}{\bibfnamefont{P.}~\bibnamefont{Pietrulewicz}},
  \bibnamefont{and} \bibinfo{author}{\bibfnamefont{D.}~\bibnamefont{Samitz}},
  \bibinfo{journal}{Phys. Rev.} \textbf{\bibinfo{volume}{D93}},
  \bibinfo{pages}{034034} (\bibinfo{year}{2016}), \eprint{1508.04323}.

\bibitem[{\citenamefont{Ablinger et~al.}(2017)\citenamefont{Ablinger,
  Blümlein, De~Freitas, Hasselhuhn, Schneider, and
  Wißbrock}}]{Ablinger:2017err}
\bibinfo{author}{\bibfnamefont{J.}~\bibnamefont{Ablinger}},
  \bibinfo{author}{\bibfnamefont{J.}~\bibnamefont{Blümlein}},
  \bibinfo{author}{\bibfnamefont{A.}~\bibnamefont{De~Freitas}},
  \bibinfo{author}{\bibfnamefont{A.}~\bibnamefont{Hasselhuhn}},
  \bibinfo{author}{\bibfnamefont{C.}~\bibnamefont{Schneider}},
  \bibnamefont{and}
  \bibinfo{author}{\bibfnamefont{F.}~\bibnamefont{Wißbrock}},
  \bibinfo{journal}{Nucl. Phys. B} \textbf{\bibinfo{volume}{921}},
  \bibinfo{pages}{585} (\bibinfo{year}{2017}), \eprint{1705.07030}.

\bibitem[{\citenamefont{Krauss and Napoletano}(2018)}]{Krauss:2017wmx}
\bibinfo{author}{\bibfnamefont{F.}~\bibnamefont{Krauss}} \bibnamefont{and}
  \bibinfo{author}{\bibfnamefont{D.}~\bibnamefont{Napoletano}},
  \bibinfo{journal}{Phys. Rev. D} \textbf{\bibinfo{volume}{98}},
  \bibinfo{pages}{096002} (\bibinfo{year}{2018}), \eprint{1712.06832}.

\bibitem[{\citenamefont{Forte et~al.}(2019)\citenamefont{Forte, Giani, and
  Napoletano}}]{Forte:2019hjc}
\bibinfo{author}{\bibfnamefont{S.}~\bibnamefont{Forte}},
  \bibinfo{author}{\bibfnamefont{T.}~\bibnamefont{Giani}}, \bibnamefont{and}
  \bibinfo{author}{\bibfnamefont{D.}~\bibnamefont{Napoletano}},
  \bibinfo{journal}{Eur. Phys. J. C} \textbf{\bibinfo{volume}{79}},
  \bibinfo{pages}{609} (\bibinfo{year}{2019}), \eprint{1905.02207}.

\bibitem[{\citenamefont{Gehrmann-De~Ridder
  et~al.}(2016)\citenamefont{Gehrmann-De~Ridder, Gehrmann, Glover, Huss, and
  Morgan}}]{Ridder:2015dxa}
\bibinfo{author}{\bibfnamefont{A.}~\bibnamefont{Gehrmann-De~Ridder}},
  \bibinfo{author}{\bibfnamefont{T.}~\bibnamefont{Gehrmann}},
  \bibinfo{author}{\bibfnamefont{E.~W.~N.} \bibnamefont{Glover}},
  \bibinfo{author}{\bibfnamefont{A.}~\bibnamefont{Huss}}, \bibnamefont{and}
  \bibinfo{author}{\bibfnamefont{T.~A.} \bibnamefont{Morgan}},
  \bibinfo{journal}{Phys. Rev. Lett.} \textbf{\bibinfo{volume}{117}},
  \bibinfo{pages}{022001} (\bibinfo{year}{2016}), \eprint{1507.02850}.

\bibitem[{\citenamefont{Gehrmann-De~Ridder
  et~al.}(2005{\natexlab{a}})\citenamefont{Gehrmann-De~Ridder, Gehrmann, and
  Glover}}]{GehrmannDeRidder:2005cm}
\bibinfo{author}{\bibfnamefont{A.}~\bibnamefont{Gehrmann-De~Ridder}},
  \bibinfo{author}{\bibfnamefont{T.}~\bibnamefont{Gehrmann}}, \bibnamefont{and}
  \bibinfo{author}{\bibfnamefont{E.~W.~N.} \bibnamefont{Glover}},
  \bibinfo{journal}{JHEP} \textbf{\bibinfo{volume}{09}}, \bibinfo{pages}{056}
  (\bibinfo{year}{2005}{\natexlab{a}}), \eprint{hep-ph/0505111}.

\bibitem[{\citenamefont{Gehrmann-De~Ridder
  et~al.}(2005{\natexlab{b}})\citenamefont{Gehrmann-De~Ridder, Gehrmann, and
  Glover}}]{GehrmannDeRidder:2005aw}
\bibinfo{author}{\bibfnamefont{A.}~\bibnamefont{Gehrmann-De~Ridder}},
  \bibinfo{author}{\bibfnamefont{T.}~\bibnamefont{Gehrmann}}, \bibnamefont{and}
  \bibinfo{author}{\bibfnamefont{E.~W.~N.} \bibnamefont{Glover}},
  \bibinfo{journal}{Phys. Lett.} \textbf{\bibinfo{volume}{B612}},
  \bibinfo{pages}{49} (\bibinfo{year}{2005}{\natexlab{b}}),
  \eprint{hep-ph/0502110}.

\bibitem[{\citenamefont{Gehrmann-De~Ridder
  et~al.}(2005{\natexlab{c}})\citenamefont{Gehrmann-De~Ridder, Gehrmann, and
  Glover}}]{GehrmannDeRidder:2005hi}
\bibinfo{author}{\bibfnamefont{A.}~\bibnamefont{Gehrmann-De~Ridder}},
  \bibinfo{author}{\bibfnamefont{T.}~\bibnamefont{Gehrmann}}, \bibnamefont{and}
  \bibinfo{author}{\bibfnamefont{E.~W.~N.} \bibnamefont{Glover}},
  \bibinfo{journal}{Phys. Lett.} \textbf{\bibinfo{volume}{B612}},
  \bibinfo{pages}{36} (\bibinfo{year}{2005}{\natexlab{c}}),
  \eprint{hep-ph/0501291}.

\bibitem[{\citenamefont{Daleo et~al.}(2007)\citenamefont{Daleo, Gehrmann, and
  Maitre}}]{Daleo:2006xa}
\bibinfo{author}{\bibfnamefont{A.}~\bibnamefont{Daleo}},
  \bibinfo{author}{\bibfnamefont{T.}~\bibnamefont{Gehrmann}}, \bibnamefont{and}
  \bibinfo{author}{\bibfnamefont{D.}~\bibnamefont{Maitre}},
  \bibinfo{journal}{JHEP} \textbf{\bibinfo{volume}{04}}, \bibinfo{pages}{016}
  (\bibinfo{year}{2007}), \eprint{hep-ph/0612257}.

\bibitem[{\citenamefont{Daleo et~al.}(2010)\citenamefont{Daleo,
  Gehrmann-De~Ridder, Gehrmann, and Luisoni}}]{Daleo:2009yj}
\bibinfo{author}{\bibfnamefont{A.}~\bibnamefont{Daleo}},
  \bibinfo{author}{\bibfnamefont{A.}~\bibnamefont{Gehrmann-De~Ridder}},
  \bibinfo{author}{\bibfnamefont{T.}~\bibnamefont{Gehrmann}}, \bibnamefont{and}
  \bibinfo{author}{\bibfnamefont{G.}~\bibnamefont{Luisoni}},
  \bibinfo{journal}{JHEP} \textbf{\bibinfo{volume}{01}}, \bibinfo{pages}{118}
  (\bibinfo{year}{2010}), \eprint{0912.0374}.

\bibitem[{\citenamefont{Boughezal et~al.}(2011)\citenamefont{Boughezal,
  Gehrmann-De~Ridder, and Ritzmann}}]{Boughezal:2010mc}
\bibinfo{author}{\bibfnamefont{R.}~\bibnamefont{Boughezal}},
  \bibinfo{author}{\bibfnamefont{A.}~\bibnamefont{Gehrmann-De~Ridder}},
  \bibnamefont{and} \bibinfo{author}{\bibfnamefont{M.}~\bibnamefont{Ritzmann}},
  \bibinfo{journal}{JHEP} \textbf{\bibinfo{volume}{02}}, \bibinfo{pages}{098}
  (\bibinfo{year}{2011}), \eprint{1011.6631}.

\bibitem[{\citenamefont{Gehrmann and Monni}(2011)}]{Gehrmann:2011wi}
\bibinfo{author}{\bibfnamefont{T.}~\bibnamefont{Gehrmann}} \bibnamefont{and}
  \bibinfo{author}{\bibfnamefont{P.~F.} \bibnamefont{Monni}},
  \bibinfo{journal}{JHEP} \textbf{\bibinfo{volume}{12}}, \bibinfo{pages}{049}
  (\bibinfo{year}{2011}), \eprint{1107.4037}.

\bibitem[{\citenamefont{Gehrmann-De~Ridder
  et~al.}(2012)\citenamefont{Gehrmann-De~Ridder, Gehrmann, and
  Ritzmann}}]{GehrmannDeRidder:2012ja}
\bibinfo{author}{\bibfnamefont{A.}~\bibnamefont{Gehrmann-De~Ridder}},
  \bibinfo{author}{\bibfnamefont{T.}~\bibnamefont{Gehrmann}}, \bibnamefont{and}
  \bibinfo{author}{\bibfnamefont{M.}~\bibnamefont{Ritzmann}},
  \bibinfo{journal}{JHEP} \textbf{\bibinfo{volume}{10}}, \bibinfo{pages}{047}
  (\bibinfo{year}{2012}), \eprint{1207.5779}.

\bibitem[{\citenamefont{Currie et~al.}(2013)\citenamefont{Currie, Glover, and
  Wells}}]{Currie:2013vh}
\bibinfo{author}{\bibfnamefont{J.}~\bibnamefont{Currie}},
  \bibinfo{author}{\bibfnamefont{E.~W.~N.} \bibnamefont{Glover}},
  \bibnamefont{and} \bibinfo{author}{\bibfnamefont{S.}~\bibnamefont{Wells}},
  \bibinfo{journal}{JHEP} \textbf{\bibinfo{volume}{04}}, \bibinfo{pages}{066}
  (\bibinfo{year}{2013}), \eprint{1301.4693}.

\bibitem[{\citenamefont{Gauld et~al.}(2019)\citenamefont{Gauld,
  Gehrmann-De~Ridder, Glover, Huss, and Majer}}]{Gauld:2019yng}
\bibinfo{author}{\bibfnamefont{R.}~\bibnamefont{Gauld}},
  \bibinfo{author}{\bibfnamefont{A.}~\bibnamefont{Gehrmann-De~Ridder}},
  \bibinfo{author}{\bibfnamefont{E.~W.~N.} \bibnamefont{Glover}},
  \bibinfo{author}{\bibfnamefont{A.}~\bibnamefont{Huss}}, \bibnamefont{and}
  \bibinfo{author}{\bibfnamefont{I.}~\bibnamefont{Majer}},
  \bibinfo{journal}{JHEP} \textbf{\bibinfo{volume}{10}}, \bibinfo{pages}{002}
  (\bibinfo{year}{2019}), \eprint{1907.05836}.

\bibitem[{\citenamefont{Febres~Cordero
  et~al.}(2008)\citenamefont{Febres~Cordero, Reina, and
  Wackeroth}}]{FebresCordero:2008ci}
\bibinfo{author}{\bibfnamefont{F.}~\bibnamefont{Febres~Cordero}},
  \bibinfo{author}{\bibfnamefont{L.}~\bibnamefont{Reina}}, \bibnamefont{and}
  \bibinfo{author}{\bibfnamefont{D.}~\bibnamefont{Wackeroth}},
  \bibinfo{journal}{Phys. Rev.} \textbf{\bibinfo{volume}{D78}},
  \bibinfo{pages}{074014} (\bibinfo{year}{2008}), \eprint{0806.0808}.

\bibitem[{\citenamefont{Febres~Cordero
  et~al.}(2009)\citenamefont{Febres~Cordero, Reina, and
  Wackeroth}}]{Cordero:2009kv}
\bibinfo{author}{\bibfnamefont{F.}~\bibnamefont{Febres~Cordero}},
  \bibinfo{author}{\bibfnamefont{L.}~\bibnamefont{Reina}}, \bibnamefont{and}
  \bibinfo{author}{\bibfnamefont{D.}~\bibnamefont{Wackeroth}},
  \bibinfo{journal}{Phys. Rev.} \textbf{\bibinfo{volume}{D80}},
  \bibinfo{pages}{034015} (\bibinfo{year}{2009}), \eprint{0906.1923}.

\bibitem[{\citenamefont{Hirschi et~al.}(2011)\citenamefont{Hirschi, Frederix,
  Frixione, Garzelli, Maltoni, and Pittau}}]{Hirschi:2011pa}
\bibinfo{author}{\bibfnamefont{V.}~\bibnamefont{Hirschi}},
  \bibinfo{author}{\bibfnamefont{R.}~\bibnamefont{Frederix}},
  \bibinfo{author}{\bibfnamefont{S.}~\bibnamefont{Frixione}},
  \bibinfo{author}{\bibfnamefont{M.~V.} \bibnamefont{Garzelli}},
  \bibinfo{author}{\bibfnamefont{F.}~\bibnamefont{Maltoni}}, \bibnamefont{and}
  \bibinfo{author}{\bibfnamefont{R.}~\bibnamefont{Pittau}},
  \bibinfo{journal}{JHEP} \textbf{\bibinfo{volume}{05}}, \bibinfo{pages}{044}
  (\bibinfo{year}{2011}), \eprint{1103.0621}.

\bibitem[{\citenamefont{van Hameren et~al.}(2009)\citenamefont{van Hameren,
  Papadopoulos, and Pittau}}]{vanHameren:2009dr}
\bibinfo{author}{\bibfnamefont{A.}~\bibnamefont{van Hameren}},
  \bibinfo{author}{\bibfnamefont{C.~G.} \bibnamefont{Papadopoulos}},
  \bibnamefont{and} \bibinfo{author}{\bibfnamefont{R.}~\bibnamefont{Pittau}},
  \bibinfo{journal}{JHEP} \textbf{\bibinfo{volume}{09}}, \bibinfo{pages}{106}
  (\bibinfo{year}{2009}), \eprint{0903.4665}.

\bibitem[{\citenamefont{van Hameren}(2011)}]{vanHameren:2010cp}
\bibinfo{author}{\bibfnamefont{A.}~\bibnamefont{van Hameren}},
  \bibinfo{journal}{Comput. Phys. Commun.} \textbf{\bibinfo{volume}{182}},
  \bibinfo{pages}{2427} (\bibinfo{year}{2011}), \eprint{1007.4716}.

\bibitem[{\citenamefont{Mastrolia et~al.}(2012)\citenamefont{Mastrolia,
  Mirabella, and Peraro}}]{Mastrolia:2012bu}
\bibinfo{author}{\bibfnamefont{P.}~\bibnamefont{Mastrolia}},
  \bibinfo{author}{\bibfnamefont{E.}~\bibnamefont{Mirabella}},
  \bibnamefont{and} \bibinfo{author}{\bibfnamefont{T.}~\bibnamefont{Peraro}},
  \bibinfo{journal}{JHEP} \textbf{\bibinfo{volume}{06}}, \bibinfo{pages}{095}
  (\bibinfo{year}{2012}), \bibinfo{note}{[Erratum: JHEP11,128(2012)]},
  \eprint{1203.0291}.

\bibitem[{\citenamefont{Peraro}(2014)}]{Peraro:2014cba}
\bibinfo{author}{\bibfnamefont{T.}~\bibnamefont{Peraro}},
  \bibinfo{journal}{Comput. Phys. Commun.} \textbf{\bibinfo{volume}{185}},
  \bibinfo{pages}{2771} (\bibinfo{year}{2014}), \eprint{1403.1229}.

\bibitem[{\citenamefont{Denner et~al.}(2017)\citenamefont{Denner, Dittmaier,
  and Hofer}}]{Denner:2016kdg}
\bibinfo{author}{\bibfnamefont{A.}~\bibnamefont{Denner}},
  \bibinfo{author}{\bibfnamefont{S.}~\bibnamefont{Dittmaier}},
  \bibnamefont{and} \bibinfo{author}{\bibfnamefont{L.}~\bibnamefont{Hofer}},
  \bibinfo{journal}{Comput. Phys. Commun.} \textbf{\bibinfo{volume}{212}},
  \bibinfo{pages}{220} (\bibinfo{year}{2017}), \eprint{1604.06792}.

\bibitem[{\citenamefont{Ball et~al.}(2017)}]{Ball:2017nwa}
\bibinfo{author}{\bibfnamefont{R.~D.} \bibnamefont{Ball}} \bibnamefont{et~al.}
  (\bibinfo{collaboration}{NNPDF}), \bibinfo{journal}{Eur. Phys. J.}
  \textbf{\bibinfo{volume}{C77}}, \bibinfo{pages}{663} (\bibinfo{year}{2017}),
  \eprint{1706.00428}.

\bibitem[{\citenamefont{Buckley et~al.}(2015)\citenamefont{Buckley, Ferrando,
  Lloyd, Nordström, Page et~al.}}]{Buckley:2014ana}
\bibinfo{author}{\bibfnamefont{A.}~\bibnamefont{Buckley}},
  \bibinfo{author}{\bibfnamefont{J.}~\bibnamefont{Ferrando}},
  \bibinfo{author}{\bibfnamefont{S.}~\bibnamefont{Lloyd}},
  \bibinfo{author}{\bibfnamefont{K.}~\bibnamefont{Nordström}},
  \bibinfo{author}{\bibfnamefont{B.}~\bibnamefont{Page}}, \bibnamefont{et~al.},
  \bibinfo{journal}{Eur.Phys.J.} \textbf{\bibinfo{volume}{C75}},
  \bibinfo{pages}{132} (\bibinfo{year}{2015}), \eprint{1412.7420}.

\bibitem[{\citenamefont{Gauld et~al.}(2017)\citenamefont{Gauld,
  Gehrmann-De~Ridder, Gehrmann, Glover, and Huss}}]{Gauld:2017tww}
\bibinfo{author}{\bibfnamefont{R.}~\bibnamefont{Gauld}},
  \bibinfo{author}{\bibfnamefont{A.}~\bibnamefont{Gehrmann-De~Ridder}},
  \bibinfo{author}{\bibfnamefont{T.}~\bibnamefont{Gehrmann}},
  \bibinfo{author}{\bibfnamefont{E.~W.~N.} \bibnamefont{Glover}},
  \bibnamefont{and} \bibinfo{author}{\bibfnamefont{A.}~\bibnamefont{Huss}},
  \bibinfo{journal}{JHEP} \textbf{\bibinfo{volume}{11}}, \bibinfo{pages}{003}
  (\bibinfo{year}{2017}), \eprint{1708.00008}.

\bibitem[{\citenamefont{Chatrchyan et~al.}(2012)}]{Chatrchyan:2012vr}
\bibinfo{author}{\bibfnamefont{S.}~\bibnamefont{Chatrchyan}}
  \bibnamefont{et~al.} (\bibinfo{collaboration}{CMS}), \bibinfo{journal}{JHEP}
  \textbf{\bibinfo{volume}{06}}, \bibinfo{pages}{126} (\bibinfo{year}{2012}),
  \eprint{1204.1643}.

\bibitem[{\citenamefont{Chatrchyan et~al.}(2013)}]{Chatrchyan:2013zja}
\bibinfo{author}{\bibfnamefont{S.}~\bibnamefont{Chatrchyan}}
  \bibnamefont{et~al.} (\bibinfo{collaboration}{CMS}), \bibinfo{journal}{JHEP}
  \textbf{\bibinfo{volume}{12}}, \bibinfo{pages}{039} (\bibinfo{year}{2013}),
  \eprint{1310.1349}.

\bibitem[{\citenamefont{Chatrchyan
  et~al.}(2014{\natexlab{a}})}]{Chatrchyan:2014dha}
\bibinfo{author}{\bibfnamefont{S.}~\bibnamefont{Chatrchyan}}
  \bibnamefont{et~al.} (\bibinfo{collaboration}{CMS}), \bibinfo{journal}{JHEP}
  \textbf{\bibinfo{volume}{06}}, \bibinfo{pages}{120}
  (\bibinfo{year}{2014}{\natexlab{a}}), \eprint{1402.1521}.

\bibitem[{\citenamefont{Aad et~al.}(2014{\natexlab{a}})}]{Aad:2014dvb}
\bibinfo{author}{\bibfnamefont{G.}~\bibnamefont{Aad}} \bibnamefont{et~al.}
  (\bibinfo{collaboration}{ATLAS}), \bibinfo{journal}{JHEP}
  \textbf{\bibinfo{volume}{10}}, \bibinfo{pages}{141}
  (\bibinfo{year}{2014}{\natexlab{a}}), \eprint{1407.3643}.

\bibitem[{\citenamefont{Sirunyan et~al.}(2020)}]{Sirunyan:2020lgh}
\bibinfo{author}{\bibfnamefont{A.~M.} \bibnamefont{Sirunyan}}
  \bibnamefont{et~al.} (\bibinfo{collaboration}{CMS}) (\bibinfo{year}{2020}),
  \eprint{2001.06899}.

\bibitem[{\citenamefont{Aad et~al.}(2020)}]{Aad:2020gfi}
\bibinfo{author}{\bibfnamefont{G.}~\bibnamefont{Aad}} \bibnamefont{et~al.}
  (\bibinfo{collaboration}{ATLAS}) (\bibinfo{year}{2020}), \eprint{2003.11960}.

\bibitem[{\citenamefont{Adye}(2011)}]{Adye:2011gm}
\bibinfo{author}{\bibfnamefont{T.}~\bibnamefont{Adye}}, in
  \emph{\bibinfo{booktitle}{{Proceedings, PHYSTAT 2011 Workshop on Statistical
  Issues Related to Discovery Claims in Search Experiments and Unfolding,
  CERN,Geneva, Switzerland 17-20 January 2011}}}, \bibinfo{organization}{CERN}
  (\bibinfo{publisher}{CERN}, \bibinfo{address}{Geneva}, \bibinfo{year}{2011}),
  pp. \bibinfo{pages}{313--318}, \eprint{1105.1160}.

\bibitem[{\citenamefont{D'Agostini}(1995)}]{DAgostini:1994fjx}
\bibinfo{author}{\bibfnamefont{G.}~\bibnamefont{D'Agostini}},
  \bibinfo{journal}{Nucl.\ Instrum.\ Meth.\ A} \textbf{\bibinfo{volume}{362}},
  \bibinfo{pages}{487} (\bibinfo{year}{1995}).

\bibitem[{\citenamefont{Sjöstrand et~al.}(2015)\citenamefont{Sjöstrand, Ask,
  Christiansen, Corke, Desai et~al.}}]{Sjostrand:2014zea}
\bibinfo{author}{\bibfnamefont{T.}~\bibnamefont{Sjöstrand}},
  \bibinfo{author}{\bibfnamefont{S.}~\bibnamefont{Ask}},
  \bibinfo{author}{\bibfnamefont{J.~R.} \bibnamefont{Christiansen}},
  \bibinfo{author}{\bibfnamefont{R.}~\bibnamefont{Corke}},
  \bibinfo{author}{\bibfnamefont{N.}~\bibnamefont{Desai}},
  \bibnamefont{et~al.}, \bibinfo{journal}{Comput.Phys.Commun.}
  \textbf{\bibinfo{volume}{191}}, \bibinfo{pages}{159} (\bibinfo{year}{2015}),
  \eprint{1410.3012}.

\bibitem[{\citenamefont{Chatrchyan
  et~al.}(2014{\natexlab{b}})}]{Chatrchyan:2013mza}
\bibinfo{author}{\bibfnamefont{S.}~\bibnamefont{Chatrchyan}}
  \bibnamefont{et~al.} (\bibinfo{collaboration}{CMS}),
  \bibinfo{journal}{Phys.Rev.} \textbf{\bibinfo{volume}{D90}},
  \bibinfo{pages}{032004} (\bibinfo{year}{2014}{\natexlab{b}}),
  \eprint{1312.6283}.

\bibitem[{\citenamefont{Aad et~al.}(2014{\natexlab{b}})}]{Aad:2014xca}
\bibinfo{author}{\bibfnamefont{G.}~\bibnamefont{Aad}} \bibnamefont{et~al.}
  (\bibinfo{collaboration}{ATLAS}), \bibinfo{journal}{JHEP}
  \textbf{\bibinfo{volume}{1405}}, \bibinfo{pages}{068}
  (\bibinfo{year}{2014}{\natexlab{b}}), \eprint{1402.6263}.

\bibitem[{\citenamefont{Alekhin et~al.}(2014)\citenamefont{Alekhin,
  Bl{\"u}mlein, Caminadac, Lipka, Lohwasser et~al.}}]{Alekhin:2014sya}
\bibinfo{author}{\bibfnamefont{S.}~\bibnamefont{Alekhin}},
  \bibinfo{author}{\bibfnamefont{J.}~\bibnamefont{Bl{\"u}mlein}},
  \bibinfo{author}{\bibfnamefont{L.}~\bibnamefont{Caminadac}},
  \bibinfo{author}{\bibfnamefont{K.}~\bibnamefont{Lipka}},
  \bibinfo{author}{\bibfnamefont{K.}~\bibnamefont{Lohwasser}},
  \bibnamefont{et~al.} (\bibinfo{year}{2014}), \eprint{1404.6469}.

\bibitem[{\citenamefont{Alekhin et~al.}(2018)\citenamefont{Alekhin,
  Bl{\"u}mlein, and Moch}}]{Alekhin:2017olj}
\bibinfo{author}{\bibfnamefont{S.}~\bibnamefont{Alekhin}},
  \bibinfo{author}{\bibfnamefont{J.}~\bibnamefont{Bl{\"u}mlein}},
  \bibnamefont{and} \bibinfo{author}{\bibfnamefont{S.}~\bibnamefont{Moch}},
  \bibinfo{journal}{Phys. Lett.} \textbf{\bibinfo{volume}{B777}},
  \bibinfo{pages}{134} (\bibinfo{year}{2018}), \eprint{1708.01067}.

\bibitem[{\citenamefont{Boettcher et~al.}(2016)\citenamefont{Boettcher, Ilten,
  and Williams}}]{Boettcher:2015sqn}
\bibinfo{author}{\bibfnamefont{T.}~\bibnamefont{Boettcher}},
  \bibinfo{author}{\bibfnamefont{P.}~\bibnamefont{Ilten}}, \bibnamefont{and}
  \bibinfo{author}{\bibfnamefont{M.}~\bibnamefont{Williams}},
  \bibinfo{journal}{Phys. Rev.} \textbf{\bibinfo{volume}{D93}},
  \bibinfo{pages}{074008} (\bibinfo{year}{2016}), \eprint{1512.06666}.

\end{thebibliography}

\end{document}